\documentclass[prl, aps,twocolumn,notitlepage,superscriptaddress,nofootinbib]{revtex4-2}

\usepackage{xcolor,graphicx}
\usepackage{amssymb,amsmath,graphicx}
\usepackage[utf8]{inputenc}
\usepackage{comment}
\usepackage{braket}
\usepackage[normalem]{ulem}
\usepackage{hyperref,url}
\usepackage{float}

\usepackage{siunitx} 
\newcolumntype{T}{j[table-format=3.5]}
\usepackage{cellspace}

\usepackage{xcolor}
\hypersetup{
    colorlinks,
    linkcolor={red!50!black},
    citecolor={blue!50!black},
    urlcolor={blue!80!black}
}

\usepackage{longtable}
\usepackage{mathtools}

\newcommand{\field}{\Hat{\mathcal{E}}}

\begin{document}

\title{Multiplexed quantum repeaters with hot multimode alkali-noble gas memories}

\author{Alexandre Barbosa}
\affiliation{Physics Department, Instituto Superior Técnico, 1049-001 Lisboa, Portugal}
\email{alexandre.barbosa@tecnico.ulisboa.pt}

\author{Hugo Ter\c cas}
\affiliation{GoLP/Instituto de Plasmas e Fusão Nuclear, Instituto Superior Técnico, 1049-001 Lisboa, Portugal}

\author{Emmanuel Zambrini Cruzeiro}
\affiliation{Instituto de Telecomunica\c c\~oes, 1049-001 Lisboa, Portugal}
\email{emmanuel.cruzeiro@lx.it.pt}

\begin{abstract}
We propose a non-cryogenic optical quantum memory for noble-gas nuclear spins based on the Atomic Frequency Comb (AFC) protocol. Owing to the hours-long coherence lifetimes of the noble-gas spins and the large bandwidth provided by the AFC independently of the optical depth, we estimate a time-bandwidth product of up to $9.7 \times 10^{15}$ for a realistic experimental configuration, using alkali-metal atoms as mediators. Leveraging this long-lived multimode memory, we propose a fiber-based quantum repeater scheme that could enable entanglement distribution across distances over $2000 \ \mathrm{km}$ with only $8$ elementary links, operating fully without cryogenics. Finally, we discuss how these quantum memories can enhance rates in satellite quantum communication networks.
\end{abstract}

\maketitle

\textit{Introduction}. Optical quantum memories are a crucial building block for quantum repeaters \cite{DLCZ, Gisin2011} enabling long-distance quantum key distribution \cite{Kimble2008} and secure direct communication \cite{Guo2017} or interferometric telescopes with arbitrarily long baselines \cite{Croke2012}. A quantum memory can simultaneously act as a single-photon source \cite{Chen2006, Dideriksen2021}, synchronize independent optical photons  \cite{Makino2016, Davidson2023} and coherently filter them \cite{Gao2019}, with applications in quantum information processing \cite{Nunn2013}, including linear optical quantum computing \cite{Kok2007}. In quantum sensing applications, quantum memories can be used to increase the acquisition time, for example enhancing sensitivity in NMR spectroscopy \cite{Zaiser2016} and spatial resolution in nuclear spin imaging for fundamental biology and drug discovery \cite{Ajoy2015}. 

The Atomic Frequency Comb (AFC) protocol \cite{Afzelius2009, deRiedmatten2008} is one of the most promising quantum memory schemes proposed to date, demonstrating high efficiency \cite{Rippe2013} over impressive bandwidths \cite{Tittel2020} with long storage times \cite{Ortu2022} and unprecedented multi-modality \cite{Seri2019} when implemented in rare-earth ion-doped crystals at cryogenic temperatures. In addition to finding applications in telecommunications, AFC quantum memories can be used for optical detection of ultrasound, enabling highly-sensitive real-time imaging of soft biological tissues \cite{McAuslan2012}.

A room-temperature AFC memory for light was recently realized in proof-of-principle experiments with cesium \cite{Ledingham2021a} and rubidium \cite{Ledingham2023} vapors, promising to extend the range of AFC applications by reducing the complexity of experimental setups as a non-cryogenic platform, thereby increasing its robustness and scalability. 

Odd isotopes of noble-gas atoms, such as $^{3} \mathrm{K}$ and $^{129} \mathrm{Xe}$, have nuclear spins exhibiting relaxation lifetimes surpassing hundreds of hours \cite{Walker2017}, corresponding to the longest-living macroscopic quantum systems presently known \cite{Firstenberg2020a}. Although optically inaccessible, they can be interfaced with light, with alkali atoms acting as mediators \cite{FirstenbergArxiv} through stochastic spin-exchange collisions \cite{FirstenbergPRX}. 

In this Letter, we propose a hot multimode optical quantum memory for noble-gas nuclear spins based on the AFC protocol and characterize its performance through numerical simulations. Owing to the exceptionally long coherence lifetime of the noble-gas spins and the intrinsic multi-modality of the AFC for time-bin encoded qubits \cite{Nunn2008}, our proposed scheme would enable long-lived storage and retrieval of multiple photonic modes at room temperature, with an estimated time-bandwidth product of up to $10^{16}$, three orders of magnitude higher than a previous proposal \cite{Simon2023} for this hybrid-gas system. 

Finally, we perform numerical estimates to show that a quantum repeater architecture that operates without cryogenics using these multimode quantum memories and current technology can distribute entanglement over long distances with high fidelity, both in fiber and free-space.

\begin{figure}[b!]
\centering
\includegraphics[width=0.9\columnwidth]{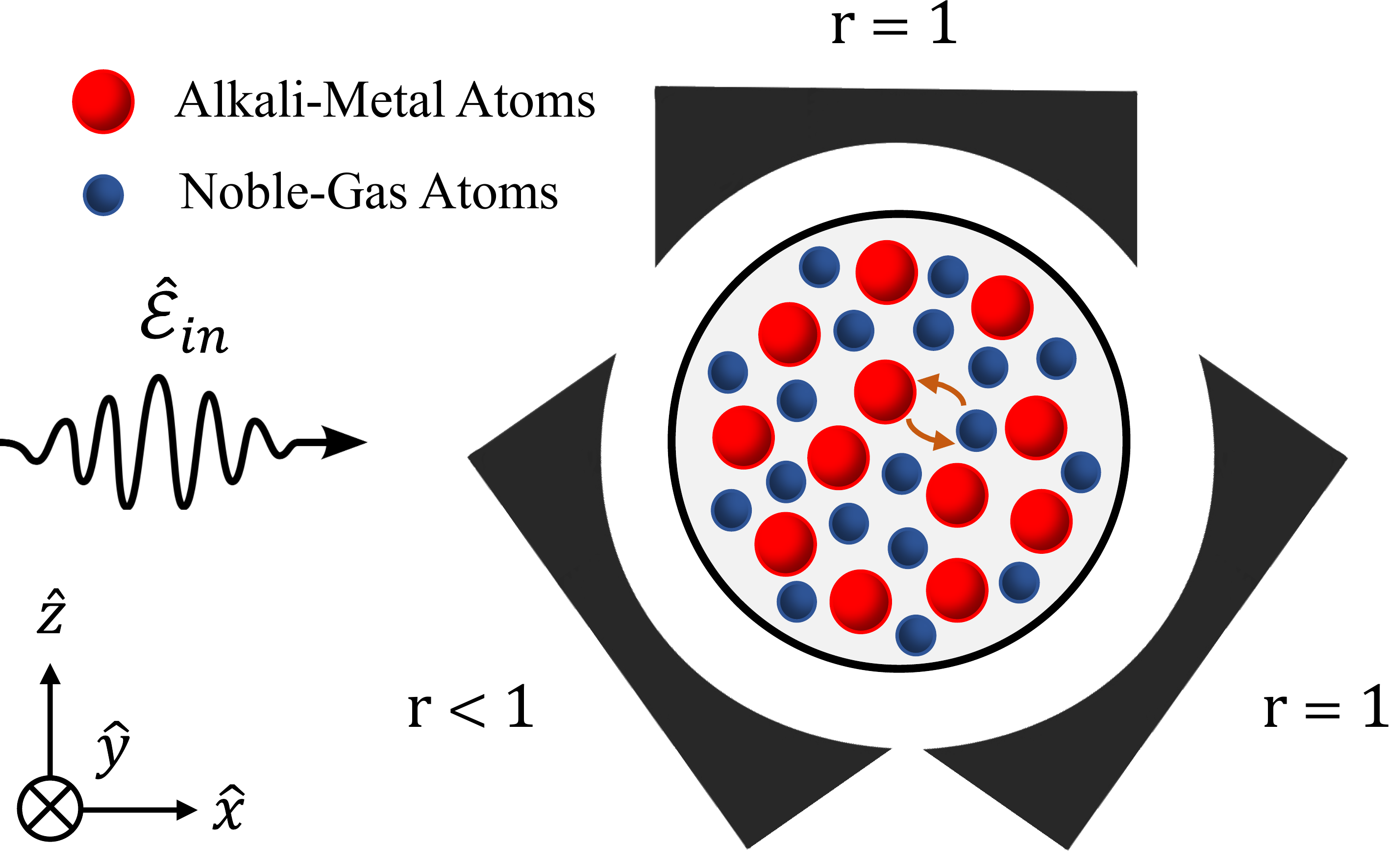}
\caption{(Color online) An optical quantum memory implemented on a mixture of alkali-metal (red) and noble-gas (blue) atoms contained in a spherical glass cell inside an optical cavity, with one partially transmitting mirror (the input-output coupler). A quantum signal field $\field_{\mathrm{in}}$ interacts with the alkali atoms, which coherently couple to the noble-gas atoms through stochastic spin-exchange collisions (orange arrows).}
\label{fig:hybrid-gas}
\end{figure}

\textit{Quantum memory protocol}. We consider an ensemble of $N_a$ alkali-metal and $N_b$ noble-gas atoms in a glass cell of volume $V$ 
inside an optical cavity with one partially transmitting mirror, as shown in Fig. \ref{fig:hybrid-gas}. Initially, the alkali spins are polarized along $\hat{z}$ through optical pumping while the noble-gas spin ensemble is in turn polarized through spin-exchange optical pumping techniques \cite{Happer1997}.

We model the alkali-metal atoms as $\Lambda$-systems, with ground state $\ket{g}$ and intermediate state $\ket{s}$ both optically connected to the excited state, $\ket{e}$, and each noble-gas atom as a two-level system with ground state $\ket{\Downarrow}$ and excited state $\ket{\Uparrow}$. A quantum light field $\field$ couples the ground and excited states, while a classical control field with Rabi frequency $\Omega$ couples to the $\ket{s}-\ket{e}$ transition.

A frequency comb can be prepared in the inhomogeneously broadened $\ket{g}-\ket{e}$ transition via piecewise adiabatic passage techniques \cite{Ahufinger2018},  velocity-selective optical pumping with narrowband frequency light modes \cite{Ledingham2021b} or an optical frequency comb from a mode-locked laser \cite{Aumiler2005}, where the laser repetition rate sets the peak separation. We suppose that an AFC with peak width (FWHM) $\gamma$, separation $\Delta$ and overall width $\Gamma$ is prepared by either of these methods, and denote its finesse by $F = \Delta/\gamma$. 

The evolution of the (slowly-varying) cavity field $\field$ is then governed by the Heisenberg-Langevin equation \cite{Afzelius2010}
\begin{equation}
    \partial_t \field = - \kappa \field + \sqrt{2 \kappa} \field_{\mathrm{in}} + i g_0^2 \wp \int_{-\infty}^{+\infty} d\delta \ n(\delta) \  \hat{\sigma}_{ge}(t; \delta),
    \label{eq:dynamics-field-cavity}
\end{equation}
where $\kappa$ is the cavity decay rate, $g_0$ is the atom-cavity coupling constant,  $\wp$ is the dipole moment for the $\ket{g}-\ket{e}$ transition, $n(\delta)$ is the atomic spectral distribution and $\hat{\sigma}_{ge}$ is the atomic polarization at detuning $\delta$, obeying
\begin{equation}
    \partial_t\hat{\sigma}_ {ge} = - (i \delta + \gamma_p ) \hat{\sigma}_{ge} + i \wp \field,
    \label{eq:dynamics-atoms-cavity}
\end{equation}
where $\gamma_p$ is the homogeneous linewidth of the optical transition. In the fast cavity limit, the incoming and outgoing cavity fields obey the input-output relation
\begin{equation}
    \field_{\mathrm{out}} - \field_{\mathrm{in}} = \sqrt{2\kappa} \field. 
    \label{eq:input-output}
\end{equation}
Accordingly, we consider the dynamics of the collective spin operators for the alkali and noble-gas ensembles, 
\begin{equation}
 \ \Hat{\mathcal{S}} (\mathbf{r}, t)  = \frac{\hat{\sigma}_{gs} (\mathbf{r}, t) }{\sqrt{p_a n_a}}, \qquad \Hat{\mathcal{K}} (\mathbf{r}, t)  = \frac{\hat{\sigma}_{\Downarrow \Uparrow} (\mathbf{r}, t) }{\sqrt{p_b n_b}}, 
\end{equation}
where $\hat{\sigma}_{\alpha \beta}$ are continuous atomic coherences \cite{Gisin2007}, $p_a, p_b$ denote the polarization degree and $n_a, n_b$ the atomic density of each ensemble \cite{FirstenbergPRA}, along with the optical dipole
\begin{equation}
    \Hat{\mathcal{P}} (\boldsymbol{r}, t) =  \frac{1}{\sqrt{p_a n_a}} \int_{-\infty}^{+\infty} d\delta \ n(\delta) \ \hat{\sigma}_{\mathrm{ge}} (\boldsymbol{r}, t; \delta). 
    \label{eq:optical-coherence-detunings}
\end{equation}
In a frame co-rotating with the quantum signal field $\field_{\mathrm{in}}$, assuming that the excited state population remains negligible, for highly polarized ensembles $p_a, p_b \approx 1$, we obtain the equations of motion for these bosonic operators \cite{FirstenbergPRA}, 
\begin{align}
    \partial_t \Hat{\mathcal{P}} &= - (\gamma_p + i \Bar{\delta})  \Hat{\mathcal{P}}  + i \Omega \Hat{\mathcal{S}  } + i G \Hat{\mathcal{E}} + \Hat{f}_{\mathcal{P}}, \label{eq:polarization-dynamics}\\ 
    \partial_t \Hat{\mathcal{S}}&= - (\gamma_s + i \delta_s - D_a \nabla^2)  \Hat{\mathcal{S}} + i \Omega^{*} \Hat{\mathcal{P}} - i J \Hat{\mathcal{K}}  + \Hat{f}_{\mathcal{S}},  \label{eq:alkali-dynamics} \\
    \partial_t \Hat{\mathcal{K}} &= - (\gamma_k + i \delta_k - D_b \nabla^2)  \Hat{\mathcal{K}}  - i J \Hat{\mathcal{S}}  + \Hat{f}_{\mathcal{K}},     \label{eq:noble-gas-dynamics} 
\end{align}
where $\Bar{\delta}$ is the average optical detuning, $\delta_s$ and $\delta_k$ are the two-photon detunings from the alkali and noble-gas spin resonances, $\gamma_p \gg \gamma_s \gg \gamma_k$ are the decoherence rates for the optical dipole, alkali and noble-gas spin coherences, $D_a$ and $D_b$ are the diffusion coefficients for the alkali and noble-gas species, $G = \sqrt{p_a n_a} g_0 \wp$ is the collective atom-field coupling frequency and $J = \zeta \sqrt{p_a p_b n_a n_b/4}$ is the coherent spin-exchange rate, where $\zeta$ is a constant \cite{FirstenbergPRX}. \newline Here, the quantum noise processes $\Hat{f}_{\mathcal{P}}$, $\Hat{f}_{\mathcal{S}}$ and $\Hat{f}_{\mathcal{K}}$ ensure the preservation of the commutation relations of the corresponding operators by introducing vacuum noise \cite{Firstenberg2020b}.

At first, with the control field turned off ($\Omega = 0$), the quantum light field is absorbed by the alkali ensemble 
and it can then be shown from equations (\ref{eq:dynamics-field-cavity})-(\ref{eq:input-output}) that 
\begin{equation}
 \field_{\mathrm{out}}  = \frac{\kappa - Z}{\kappa + Z} \  \field_{\mathrm{in}}.
\end{equation}
where $Z = N_a g_0^2 \wp^2/\Gamma$ is the absorption rate of the cavity field by the atomic ensemble. If the cavity absorption and decay rates match, $\kappa = Z$, the input field is fully absorbed, $\field_{\mathrm{out}}= 0$ \cite{Afzelius2010}. Notably, this impedance matching condition corresponds to unit optical cooperativity, $C = N_a G^2 / \kappa \Gamma = 1$, a modest experimental constraint.

In the second stage of the protocol, a chirped adiabatic pulse is used to transfer the optical coherence onto the alkali spin-wave $\Hat{\mathcal{S}}$. A complex hyperbolic secant \cite{Minar2010} or an extended hyperbolic-square-hyperbolic (HSH) \cite{Tian2011} pulse with duration $T$ can be employed to achieve efficient broadband population transfer, such that
\begin{equation}
    \langle \Hat{\mathcal{S}}^{\dagger} \Hat{\mathcal{S}} \rangle_{T} = \left[ 1 - \mathrm{exp} \left(- \frac{\pi T \Omega^2}{(2) \Gamma}\right) \right] \langle \Hat{\mathcal{P}}^{\dagger} \Hat{\mathcal{P}} \rangle_{0}.
    \label{eq:afc-spin-wave-transfer-efficiency}
\end{equation}
with the prefactor in parentheses for the HSH pulse \cite{Businger2020}.

Then, the collective alkali excitation is coherently transferred to the noble-gas ensemble through stochastic spin-exchange collisions \cite{FirstenbergPRX} after $T' \approx  (\pi J - \gamma_s)/(2 J^2)$, 
\begin{equation}
    \langle \Hat{\mathcal{K}}^{\dagger} \Hat{\mathcal{K}} \rangle_{T'} = \mathrm{exp} \left(- \frac{\pi \gamma_s}{2 J} \right) \langle \Hat{\mathcal{S}}^{\dagger} \Hat{\mathcal{S}} \rangle_{T}.
    \label{eq:afc-noblegas-transfer-efficiency}
\end{equation}
A strong magnetic field ($\delta_k \gg J$) is subsequently applied to effectively decouple the ensembles. In the dark, the noble-gas spins function as a very long-lived quantum memory, whose storage time is bounded only by $1/\gamma_k$, potentially exceeding hundreds of hours \cite{Firstenberg2020a, FirstenbergPRA}. Finally, the magnetic field is switched off and the storage sequence $\Hat{\mathcal{P}} \to \Hat{\mathcal{S}} \to \Hat{\mathcal{K}}$ is reversed. At a time $2\pi/\Delta$ after the reversal is concluded, the periodic structure of the comb leads to a coherent re-emission of the stored pulse, 
\begin{equation}
    \field_{\mathrm{out}}(t) = - \sqrt{\eta_m} \ \field_{\mathrm{in}} \left( t - 2T' - 2T - \frac{2 \pi}{\Delta} \right), 
\end{equation}
with the memory efficiency $\eta_m$ approximately given by
\begin{equation}
    \eta_m = \left[ 1 - \mathrm{exp} \left(- \frac{\pi^2 T \Omega^2}{\Gamma}\right) \right]^2 \mathrm{exp} \left( - \frac{\pi \gamma_s}{J}\right) \mathrm{sinc}^2 \left(\frac{\pi}{F}\right) 
    \label{eq:memory-efficiency}
\end{equation}
for a frequency comb with square-shaped peaks \cite{Chaneliere2010}. 

\begin{figure*}
    \includegraphics[width=0.49\textwidth]{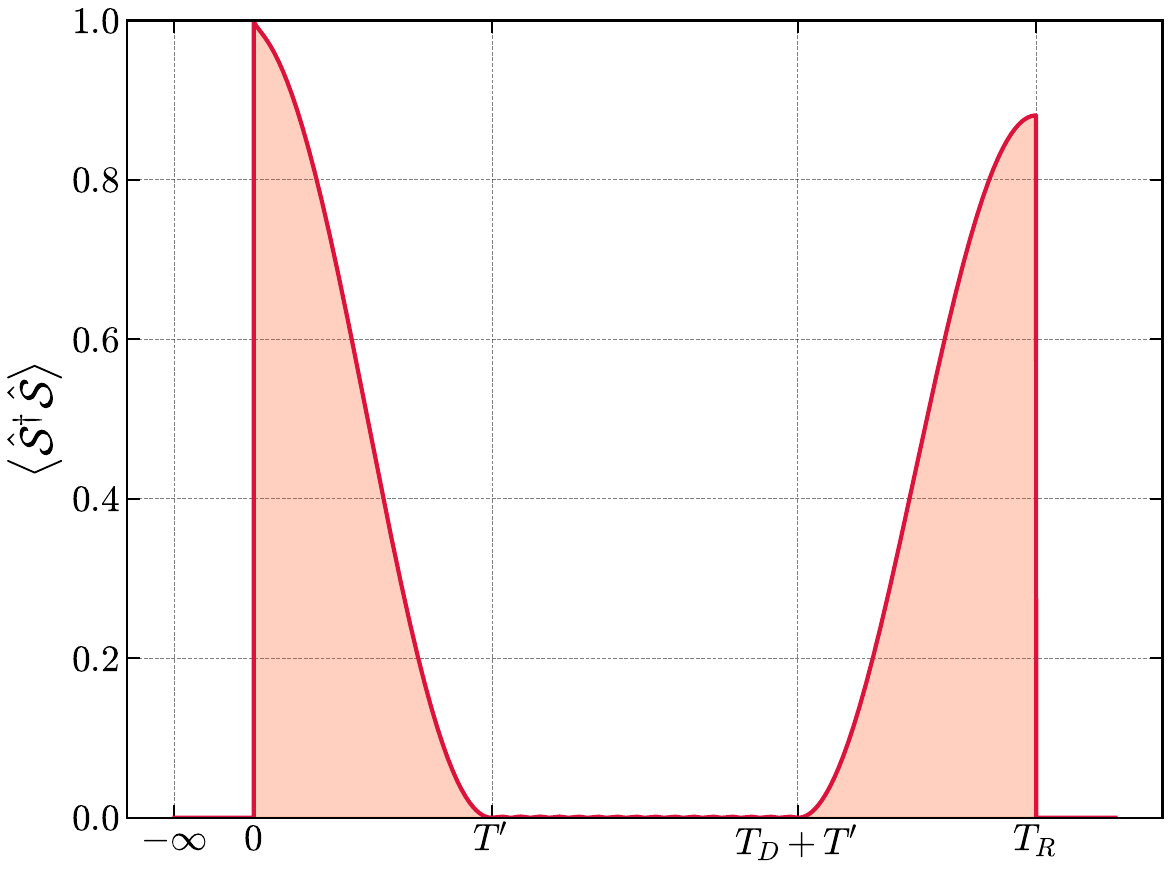}
    \includegraphics[width=0.49\textwidth]{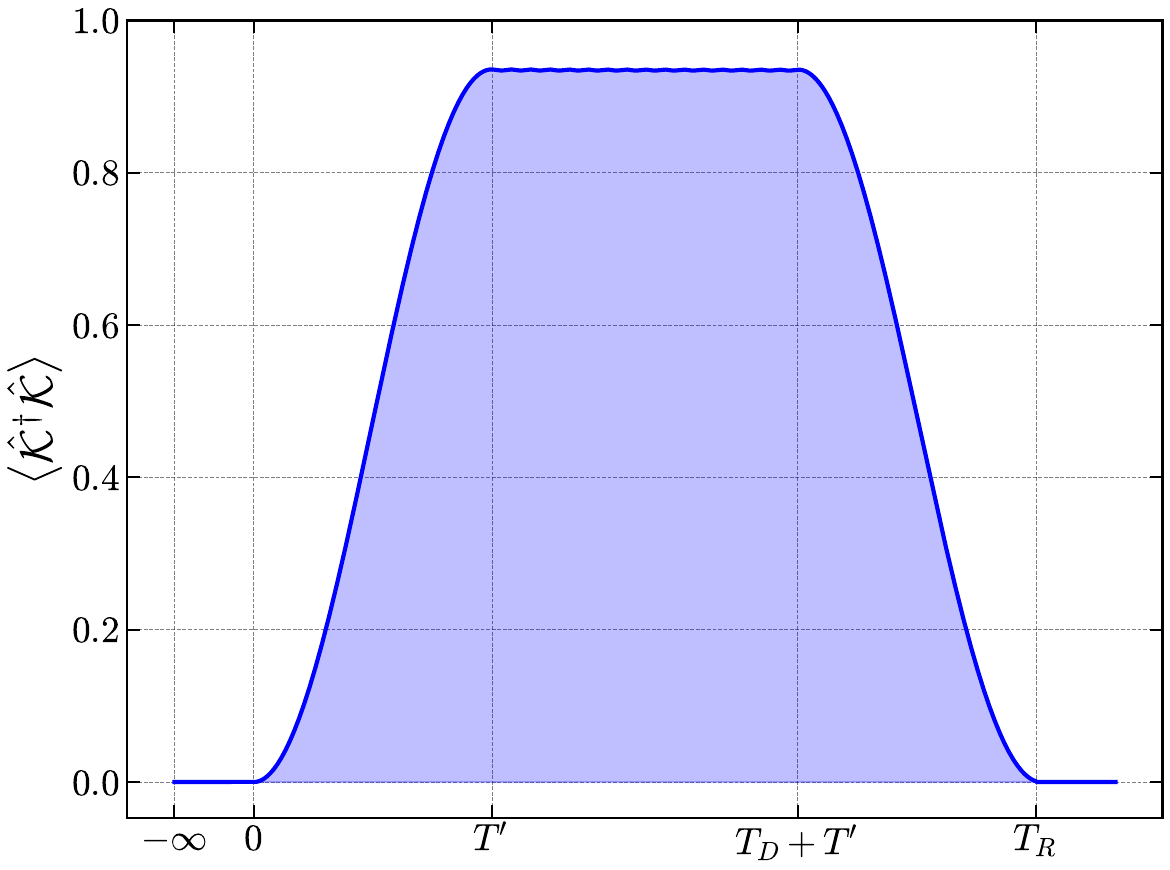}
    \caption{(Color online) Expectation values of the alkali $\langle \Hat{\mathcal{S}}^\dagger \Hat{\mathcal{S}} \rangle $ (left) and noble-gas $\langle \Hat{\mathcal{K}}^\dagger \Hat{\mathcal{K}} \rangle $ (right) spin populations throughout the protocol, with $D_a = 0.35 \ \mathrm{cm}^2\mathrm{Hz}$, $D_b = 0.70 \ \mathrm{cm}^2\mathrm{Hz}$ \cite{Firstenberg2020b}, $\delta_k = 50 \ \mathrm{kHz}$ and the parameters described in the main text.} 
    \label{fig:numerical-simulations}
\end{figure*}

\textit{Implementation}. We consider a $^3\mathrm{He-K}$ mixture confined in an uncoated spherical glass cell with radius $1 \ \mathrm{cm}$ at $215 ^\circ \mathrm{C}$, in a high-pressure configuration proposed in Ref. \cite{FirstenbergPRX} wherein we assume a $^{39}\mathrm{K}$ density of $n_a = 3 \times 10^{14} \ \mathrm{cm}^{-3}$ (at vapor pressure), a $^3\mathrm{He}$ density of $n_b = 2 \times 10^{20} \ \mathrm{cm}^{-3}$, obtaining a coherent spin-exchange rate of $J = 1000 \ \mathrm{Hz}$ for $p_a = 0.95$ and $p_b = 0.75$ \cite{Chen2014}. A buffer gas ($\mathrm{N}_2$) is introduced for quenching, at a pressure of $30 \ \mathrm{Torr}$, broadening the potassium $\mathrm{D}_1$ linewidth to $27 \ \mathrm{GHz}$. In practice, the AFC peak width $\gamma$ is limited by the optical coherence lifetime, bounded by twice the natural linewidth of the transition, $\gamma_p = 5.96 \times 2 \pi \ \mathrm{MHz}$. An intrinsic AFC efficiency of $95\%$ can be achieved for a finesse $F = 8$, corresponding to a peak separation $\Delta = 96 \ \mathrm{MHz}$. Under the assumption that the AFC bandwidth matches the broadened transition linewidth, we obtain an upper bound on the multimode capacity of $M = 2\Gamma/5\Delta = 112$ \cite{Ortu2022}. In this configuration, the alkali relaxation rate is estimated to be $\gamma_s = 17.5 \ \mathrm{Hz}$ \cite{FirstenbergPRX}. A population transfer efficiency of $98 \%$ can be attained using HSH pulses with $\pi^2 T \Omega^2/\Gamma = 4$ \cite{Ortu2022}. We therefore estimate a memory efficiency of $\eta_m = 88 \%$ using equation (\ref{eq:memory-efficiency}), mainly limited by the spin-exchange efficiency.
\begin{figure}[t]
    \includegraphics[width=\columnwidth]
    {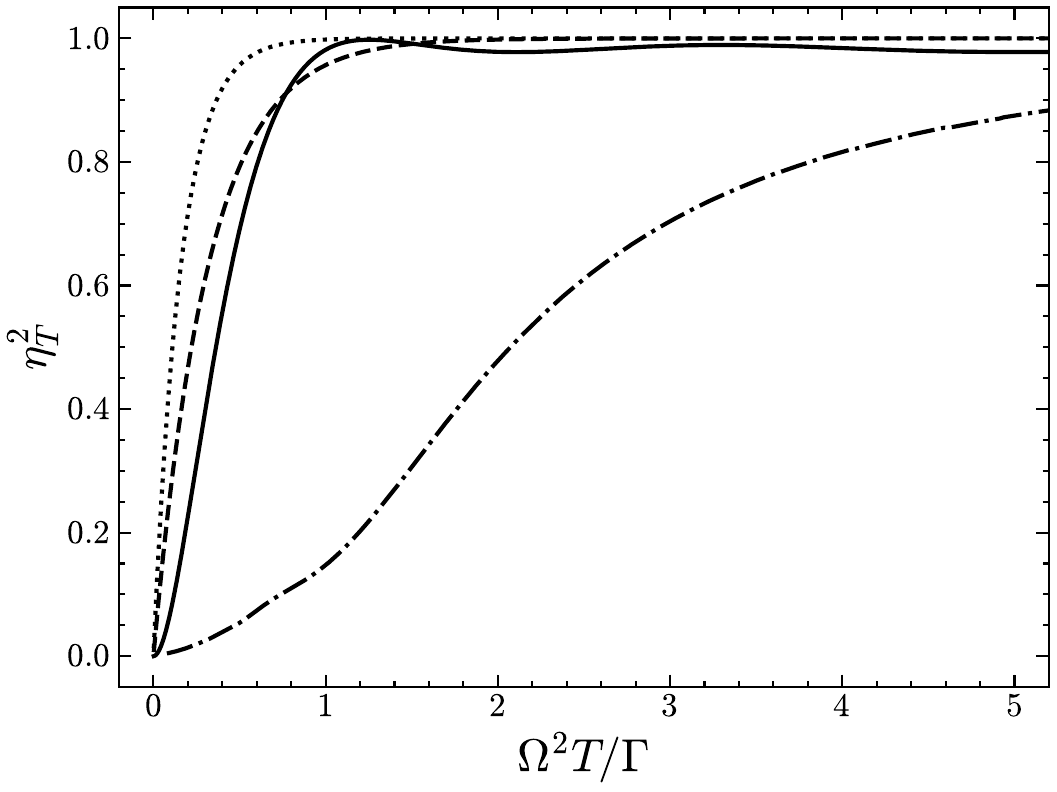}
    \caption{Transfer efficiency using hyperbolic secant pulses with optimal multimode capacity, corresponding to the minimum duration $T = 4/\Gamma$ (solid line), compared to analytical estimates (dashed line), square $\pi$-pulses (dot-dashed line) and HSH pulses (dotted line) from Eq.(\ref{eq:afc-spin-wave-transfer-efficiency}) as a function of $\Omega^2T/\Gamma$.
    }
    \label{fig:transfer-efficiency}
\end{figure}

In Fig.~\ref{fig:numerical-simulations}, we represent the expectation values of the collective spins of the alkali and noble-gas ensembles throughout the memory protocol, obtained through the numerical solution of equations (\ref{eq:alkali-dynamics})-(\ref{eq:noble-gas-dynamics}), using the method of lines for grid discretization and a time-adaptive Runge-Kutta method for numerical integration \cite{Zwicker2020}. We assume respectively destructive (Dirichlet) and non-destructive (von Neumman) boundary conditions for the alkali and noble-gas spins, representing the interaction with the surface of the glass cell \cite{Firstenberg2020b}. In Fig.~\ref{fig:transfer-efficiency}, the population transfer efficiency (back and forth) is shown for different pulse shapes, in good agreement with theoretical calculations. 

\begin{figure}[t]
    \includegraphics[width=\columnwidth]{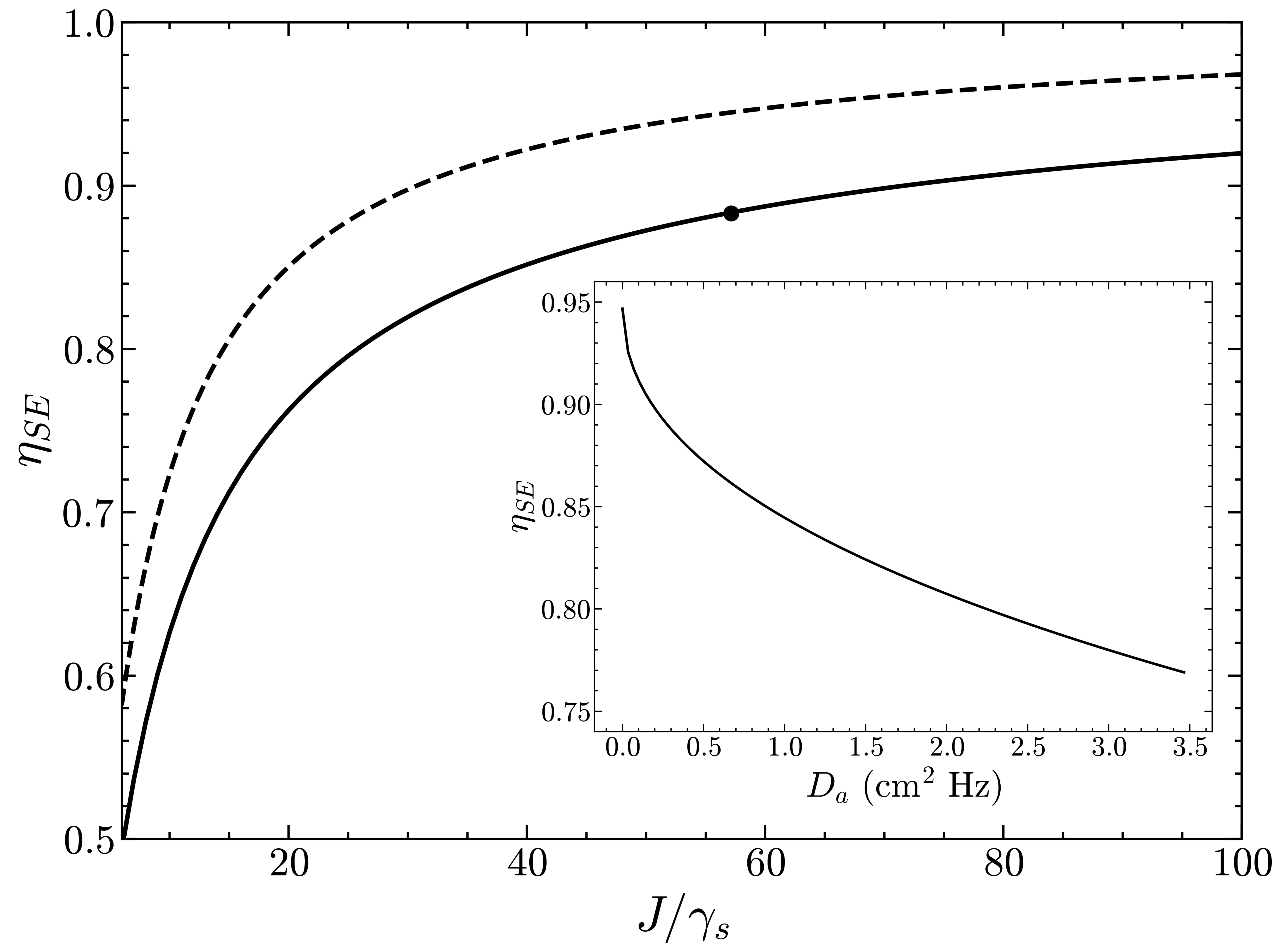}
    \caption{Efficiency of the alkali-noble gas spin exchange $\eta_{SE}$ as a function of the coupling $J/\gamma_s$ from equation (\ref{eq:afc-noblegas-transfer-efficiency}) (dashed line) and numerical simulations (solid line), reduced due to spatial mode mismatch. Inset: spin exchange efficiency as a function of the alkali diffusion rate $D_a$ for fixed $J$ (highlighted in black) in an uncoated spherical cell with radius $R = 1 \ \mathrm{cm}$. } 
    \label{fig:spin-exchange-efficiencies}
\end{figure}

We estimate a total memory efficiency of $79\%$ from our numerical solutions, compared to an analytical prediction of $88\%$ from Eq.(\ref{eq:memory-efficiency}), which is accurate only for anti-relaxation coated cells \cite{FirstenbergPRA}. In uncoated cells, the collective spin operators are subject to distinct boundary conditions so that their spatial modes only partially overlap; this mode mismatch induced by diffusion further reduces the spin-exchange efficiency, as shown in Fig. \ref{fig:spin-exchange-efficiencies}.

Anti-relaxation coatings such as alkene compounds  \cite{Balabas2010} or octadecyltrichlorosilane \cite{Seltzer2009} can improve the spin-exchange efficiency but limit the memory operation temperature \cite{FirstenbergPRA}, which sets the alkali density and therefore the coherent spin-exchange rate $J$. Although these coatings are incompatible with the high temperatures our proposed implementation requires, realizing lower-temperature configurations in coated cells or more resistant coatings may provide for higher memory efficiencies.

\textit{Quantum repeater architecture}. We consider a quantum repeater with photon pair sources and our proposed temporally multimode memories, using alkali and noble-gas atoms confined in a vapor cell. A photon pair source can also be realized using warm vapor cells \cite{Park2019, Kim2022} or parametric down-conversion in nanophotonic chips \cite{Guo2016}. 

As a first step, to generate entanglement between two nodes $A$ and $B$, the pair sources at both locations are coherently and simultaneously excited, such that each of them has a probability $p/2 \ll 1$ of creating a pair. Then, for each pair, one mode is stored in the quantum memory while the other is coupled into an optical fiber and combined in a beam splitter \cite{Gisin2011, Simon2007}. Hence, the detection of a photon after the beam splitter heralds the storage of a single photon, projecting the system into an entangled state with a single delocalized atomic excitation  \cite{DLCZ}, 
\begin{equation}
    \ket{\Phi_{AB}} = \frac{1}{\sqrt{2}} \left( \ket{1}_A \ket{0}_B + \mathrm{e}^{i \phi} \ket{0}_A \ket{1}_B \right).
\end{equation}
Using multimode quantum memories, entanglement creation can be attempted multiple times in each clock interval $L_0/c$, thus increasing the success probability $P$ to $1 - (1- P)^N \approx NP$, for $N$ distinct temporal modes \cite{Simon2007}.
\begin{figure}[b]
\centering
{
\includegraphics[width=\columnwidth]{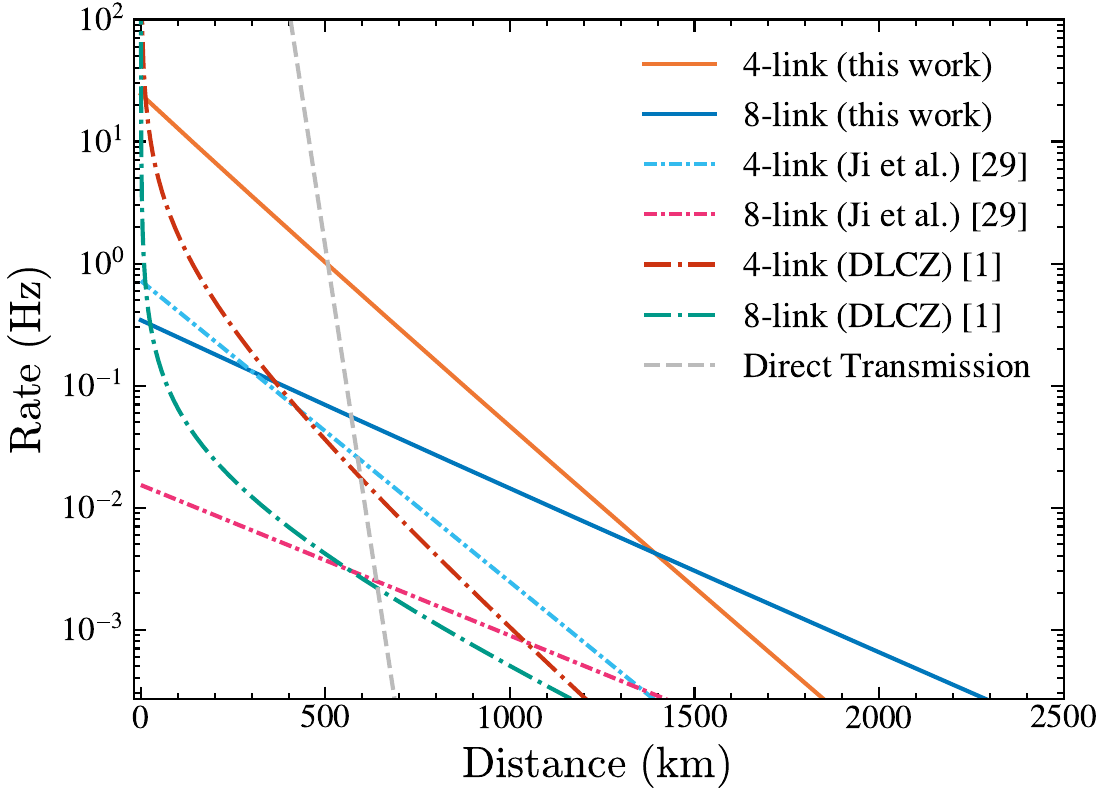}
}
\caption{(Color online) Entanglement distribution rates for a fiber-based quantum repeater as a function of distance for 4 and 8 links employing our proposed scheme (solid lines), the single-photon protocol of Ji \textit{et al.} \cite{Simon2023} (short-dashed lines), the standard DLCZ protocol \cite{DLCZ} (long-dashed lines) and direct transmission with a source rate of $10 \ \mathrm{GHz}$ (dashed line), using the parameters described in the main text for all curves.}
\label{fig:entanglement-distribution-rates}
\end{figure}
Finally, the entanglement can be teleported to a distant node by successive entanglement swapping. The total time required for entanglement distribution is \cite{Gisin2011}
\begin{equation}
    T_{\mathrm{tot}} = \left( \frac{L_0}{c} + \frac{\pi}{2J} \right) \frac{3^{n+1}}{N p \eta_c \eta_d \eta_t \eta^{n+2}} \ \prod_{k=1}^n \left(2^k (2^k-1) \eta \right), 
    \label{eq:total-time}
\end{equation}
where $L_0$ is the elementary link length, $n = \log_2 M$ is the nesting level of a repeater with $M$ nodes, $\eta_d$ is the detection efficiency, $\eta_c$ is the telecommunication frequency conversion efficiency, $\eta = \eta_m \eta_d$, $\eta_t = \mathrm{e}^{-L_0/(2 L_{\mathrm{att}})}$ is the transmission efficiency, $L_{\mathrm{att}} = 22 \ \mathrm{km}$ is the fiber attenuation length and $c = 2 \times 10^{8} \ \mathrm{m/s}$ is the speed of light in the optical fiber. We consider using silicon single-photon avalanche diodes with detection efficiency $\eta_d = 0.75$ \cite{Fang2020} and lithium niobate waveguide chips for frequency conversion \cite{Zheng2020} assuming $\eta_c = 0.8$, as proposed in Ref. \cite{Simon2023}.

A wafer of microfabricated vapor cells could provide for hundreds of independent memories in each location, speeding up entanglement distribution. An optical quantum memory in a microfabricated vapor cell has been recently demonstrated \cite{Treutlein2023}, showing that significant parallelization can be achieved with near-term technology.

In Fig. \ref{fig:entanglement-distribution-rates}, we plot the entanglement distribution rates for our proposed quantum repeater architecture with $100$ memories at each node, comparing it with the single-photon proposal of Ref. \cite{Simon2023} and the well-known DLCZ \cite{DLCZ} protocol for atomic ensembles. 
Although the long storage and retrieval times $\pi/2J \gtrsim L_0/c$ due to the slow interface between the alkali and noble-gas spins considerably reduce the entanglement distribution rates, the temporal multiplexing enabled by the AFC can easily compensate for this limitation, as Fig. \ref{fig:entanglement-distribution-rates} shows, clearly surpassing the original DLCZ protocol \cite{DLCZ} rates for a fast memory. We find that our proposed protocol outperforms direct transmission for distances over $507 \ \mathrm{km}$ and offers higher rates than the single-photon architecture proposed in Ref. \cite{Simon2023} for the hybrid alkali noble-gas memories we consider. A $4$-link repeater is optimal for distances under $1400 \ \mathrm{km}$, whereas an $8$-link repeater can achieve better rates for longer distances. 
Ultimately, the achievable distance is limited by the memory storage times, since quantum repeater protocols require it to be comparable to $T_{\rm tot}$ so that post-selection of the entangled pair is possible \cite{Gisin2011}. In the configuration we considered, the collective $^3\mathrm{He}$ nuclear spin can remain coherent for $100$ hours \cite{FirstenbergPRX}, enabling entanglement distribution for distances up to $2280 \ \mathrm{km}$. In practice, due to fiber losses this is believed to be close to the limit for terrestrial quantum networks without error correction \cite{Simon2017}, which requires considerably more complex resources. 

\emph{Conclusion}. In this Letter, we have proposed a multimode optical quantum memory based on atomic frequency combs in hot alkali-noble-gas vapors. We have shown through both analytical and numerical calculations that this memory can be operated at high efficiency for achievable physical parameters, even when accounting for the effects of spatial diffusion. With our proposed memory scheme, a quantum repeater that outperforms direct transmission could be demonstrated with only $4$ links, without cryogenic components. Overcoming this barrier would constitute an important technological milestone for ground-based quantum repeaters \cite{Simon2017}.

A solution to overcome fiber losses and achieve even longer distances is to use satellite-borne photon sources \cite{Pan2017} or quantum memories \cite{Gundogan2021}, since photon absorption and decoherence in empty space are negligible. An entangled photon pair has been distributed to locations separated by $1200 \ \mathrm{km}$ in a space-to-ground two-downlink channel \cite{Yin2017}, with losses as low as $64 \ \mathrm{dB}$. However, satellite links are plagued by intermittency due to weather and satellite-orbit shadow time, which considerably reduce entanglement distribution rates. Hence, very long-lived quantum memories such as we propose here can alleviate these issues, storing entangled pairs when the satellite link is available \cite{Simon2017}, further increasing rates due to the temporal multiplexing enabled by the AFC. In this way, our proposed memory scheme could be incorporated in hybrid or satellite-based repeater schemes \cite{Gundogan2021}, thereby contributing towards the longstanding goal of building a quantum network spanning truly global distances.

\emph{Code Availability}. The code that supports the findings of this Letter is openly available \cite{SupportingCode}.

 \emph{Acknowledgements}. The authors acknowledge Funda\c c\~ao para a Ci\^encia e a Tecnologia (FCT-Portugal) through national funds and, when applicable, co-funding by EU funds under the Projects No. UIDB/50008/2020, PTDC/FIS-OUT/3882/2020, and Contract Nos. CEECIND/00401/2018 and CEECIND/CP1653/CT0002.

\appendix

\bibliography{bibl}

\end{document}